\documentclass[twocolumn]{aa}
\usepackage{graphicx}
\usepackage{txfonts}

\usepackage{savesym}
\savesymbol{iint} \savesymbol{iiint} \savesymbol{iiiint}
\savesymbol{idotsint}
\usepackage{amsmath}
\usepackage{amsfonts}%
\usepackage{amssymb}%

\begin{document}
   \title{New near-infrared observations and lens-model constraints for \object{UM673}\thanks{Based in part on data collected at Subaru Telescope,
   which is operated by the National Astronomical Observatory of Japan.},\thanks{Reduced Subaru images presented in Figures 1 and 2 are available
   as fits files at the CDS via anonymous ftp to cdsarc.u-strasbg.fr (130.79.128.5)
or via http://cdsweb.u-strasbg.fr/cgi-bin/qcat?J/A+A/}}
   \titlerunning{NIR observations and lens model for \object{UM673}}
   \authorrunning{E.~Koptelova et al.}
       \author{E.~Koptelova\inst{\ref{inst1}}$^{,2,3}$\thanks{E-mail:
ekaterina@phys.ntu.edu.tw (EK)}\and
T.~Chiueh\inst{\ref{inst1}}\thanks{chiuehth@phys.ntu.edu.tw
(TC)}\and W.
P.~Chen\inst{\ref{inst2}}\thanks{wchen@astro.ncu.edu.tw (WPC)}\and
H. H.~Chan\inst{\ref{inst1}}}

   \offprints{E. Koptelova}

   \institute{Department of Physics, National Taiwan University, No.1, Sec. 4, Roosevelt Rd., 106 Taipei, Taiwan \label{inst1}\and Graduate Institute of Astronomy, National Central University, Jhongli City, Taoyuan County 320, Taiwan\label{inst2}\and  Sternberg Astronomical Institute (SAI), Moscow M.V. Lomonosov State University,
              Universitetskii pr. 13, 119992 Moscow, Russia\label{inst3}   }

 \date{Received July 9, 2013; accepted April 1, 2014}

\abstract {}{We performed a detailed photometric analysis of the
lensed system \object{UM673} (\object{Q0142-100}) and an analysis
of the tentative lens models.} {High-resolution adaptive optics
images of \object{UM673} taken with the Subaru telescope in the H
band were examined. We also analysed the J, H and K-band
observational data of \object{UM673} obtained with the 1.3m
telescope at the CTIO observatory.} {We present photometry of
quasar components A and B of \object{UM673}, the lens, and the
nearby bright galaxy using H-band observational data obtained with
the Subaru telescope. Based on the CTIO observations of UM673, we
also present J- and H-band photometry and estimates of the J, H
and K-band flux ratios between the two UM673 components in recent
epochs. The near-infrared fluxes of the A and B components of
\object{UM673} and their published optical fluxes are analysed to
measure extinction properties of the lensing galaxy. We estimate
the extinction-corrected flux ratio between components A and B to
be about $2.14$~mag. We discuss lens models for the \object{UM673}
system constrained with the positions of the \object{UM673}
components, their flux ratio, and the previously measured time
delay.}{}


   \maketitle

\section{Introduction}

Foreground galaxies acting as strong lenses provide important
information on the mass distribution on small scales. The mass can
be constrained by relative positions, fluxes, and time delays
between different components (images) of lensed quasars. When the
lens mass model is known, the Hubble constant can be inferred from
the time delays between the quasar images
(Refsdal~\cite{refsdal1964}). In many cases, however, the mass
model of the lens is more complex than that of a single object.
Keeton et al.~(\cite{keeton2000}) predicted that at least 25\% of
lenses lie in environment (groups and clusters) that perturbs the
lensing potential. The results of Treu et al.~(\cite{treu2008}),
who found that approximately 20\% of the lenses from their Sloan
ACS Survey belong to known groups or clusters, agree well with the
theoretical expectations. The galaxy groups are found in the
observations of many gravitationally lensed quasars (see, e.g.,
Fassnacht \& Lubin~\cite{fassnacht2002}; Faure et
al.~\cite{faure2004}; Momcheva et al.~\cite{momcheva2006};
Williams et al.~\cite{williams2006}). Often, perturbing objects
are detected on the line of sight towards lensed quasars (see,
e.g., Tonry \& Kochanek~\cite{tonry2000}; Faure et
al.~\cite{faure2004}; Fassnacht et al.~\cite{fassnacht2006};
Momcheva et al.~\cite{momcheva2006}). This line-of-sight mass
might lead to an overestimation of the mass of the main lens (see,
e.g., Wambsganss et al.~\cite{wambsganss2005}) and might have a
substantial effect on lensing properties
(Keeton~\cite{keeton2003}).

Measured time delays and flux ratios between quasar images can
help to assess properties of the perturbing mass and reduce
uncertainties in the lens model (see, e.g.,
Keeton~\cite{keeton2003}; Keeton \& Moustakas~\cite{keeton2008}).
Thus, MacLeod et al.~(\cite{macleod2009}) have identified a
secondary lensing galaxy through its effect on the fluxes of the
lensed quasar \object{H1413+117}. (This second lens is associated
with a high-redshift line-of-sight structure towards the quasar
(Kneib et al.~\cite{kneib1998}).) For the same lensed system,
Goicoechea \& Shalyapin~(\cite{goicoechea2010}) used time-delay
constraints to improve the model of the lens and to estimate the
unknown redshift of the main lensing galaxy.

In this study we focus on the lensed system \object{UM673}
(\object{Q0142-100}) discovered by Surdej et
al.~(\cite{surdej1987, surdej1988}). The system consists of images
A and B (separated by 2.2\arcsec) of a distant quasar at redshift
$z_\mathrm{q}$ = 2.719 gravitationally lensed by an elliptical
galaxy at redshift $z_\mathrm{l}$ = 0.49 (Surdej et al.
\cite{surdej1988}; Smette et al. \cite{smette1992}; Eigenbrod et
al. \cite{eigenbrod2007}). The fluxes of the two \object{UM673}
images are known to be very different, with image A being about
seven times brighter than image B. The time delay between images A
and B of \object{UM673} measured in Koptelova et
al.~(\cite{koptelova2012}) and most recently in Oscoz et
al.~(\cite{oscoz2013}) ($89\pm11$ and $72\pm22$~days,
respectively) is shorter than expected from the singular
isothermal ellipsoid model discussed in Keeton et
al.~(\cite{keeton1998}) and Leh$\rm \acute{a}$r et
al.~(\cite{lehar2000}). Leh$\rm \acute{a}$r et
al.~(\cite{lehar2000}) also estimated local tidal perturbations
produced by nearby galaxies detected in the HST images of
\object{UM673}. From their analysis, Keeton et
al.~(\cite{keeton1998}) and Leh$\rm \acute{a}$r et
al.~(\cite{lehar2000}) concluded that the lens model requires an
external shear to explain the positions of the \object{UM673}
images and their flux ratio. However, the orientation of the shear
from modelling does not coincide with the direction to any of the
nearby galaxies. Hence, new detailed observations of
\object{UM673} can be important for further understanding of the
lens mass model.

The spectroscopic observations of \object{UM673} show numerous
mass concentrations on the line of sight to the quasar. Surdej et
al.~(\cite{surdej1988}) found the same absorbtion-line systems in
the spectra of both components A and B (see also Smette et
al.~\cite{smette1992} and Rauch et al.~\cite{rauch2001}). Most
recently, Cooke et al.~(\cite{cooke2010}) discovered a damped
Ly$\alpha$ absorption (DLA) system at $z=1.63$ in the spectrum of
component A of \object{UM673}. They also found a weak Ly$\alpha$
emission line in the spectrum of component B at the redshift of
the DLA absorber.

The DLA absorber towards \object{UM673} might indicate the
existence of a massive object, a galaxy, as DLA systems are
usually associated with the progenitors of normal spiral galaxies
(Wolfe et al.~\cite{wolfe1986}). Direct observations of the DLA
galaxy could help to improve the lens model of \object{UM673}.
Unfortunately, imaging of the DLA-associated galaxies is often
complicated by their faintness and proximity to much brighter
background quasars (see, e.g., M$\o$ller et al.
~\cite{moller2002}). As demonstrated by Chun et
al.~(\cite{chun2006,chun2010}) and P$\rm \acute{e}$roux et
al.~(\cite{peroux2011}), high-resolution deep adaptive optics (AO)
observations offer a promising technique to achieve the necessary
angular resolution and sensitivity to identify faint objects at
small impact parameters from the line of sight to the bright
sources.

With the goal of studying the \object{UM673} close environment we
conduct high-resolution AO imaging of the lensed system and nearby
objects using the Subaru telescope. To be able to detect any faint
objects near the quasar, the Subaru resolution and sensitivity is
combined with the technique that can effectively remove the light
from the bright quasar images. We also present new observations of
the lensed system with the 1.3m telescope at the CTIO observatory
in the J, H, and K bands. Based on these NIR and our previous
optical observations of \object{UM673}, we obtain new estimates
for some of the observational constraints. Thus, we analyse the
extinction properties of the lensing galaxy and present new
measurements of the extinction-corrected flux ratio between the A
and B components of \object{UM673}. We discuss possible models of
the lens constrained by the measured observables (the image
positions, flux ratio, and time delay). Unlike previous studies of
\object{UM673}, we use the time delay to constrain the model of
the lens. The nearby lens perturber, if any, is likely to be a
very faint object that is difficult to detect with high
significance. Therefore, when analysing the Subaru data, the
time-delay information is incorporated to search for the probable
location of the perturber.

The paper is organised as follows: the results of the photometric
analysis of the Subaru and CTIO data are given in
Sects.~\ref{subaru} and~\ref{ctio_phot}. In Sect.~\ref{extinction}
we examine extinction properties of the lensing galaxy. The model
of the lens that can reproduce the positions of the \object{UM673}
components, their flux ratio, and the time delay is discussed in
Sect.~\ref{model}. We conclude with a summary and discussion given
in Sect.~\ref{discussion}.

\section{Observations} \label{observations}

The Subaru observations were carried out with the Infra-red Camera
and Spectrograph (IRCS) (Kobayashi et al. \cite{kobayashi2000})
using an adaptive optics system with 188 control elements (AO188)
(Hayan et al. \cite{hayano2010}) in the natural guide star (NGS)
mode. \object{UM673} was observed in the H band on February 1,
2012. The images were taken with a $1024\times1024$ ALADDIN III
CCD. The pixel scale was 20~$\mu$as, resulting in a field of view
of 21\arcsec. The gain of the CCD was 5.6 $\rm{e^{-}/ADU}$.
Imaging was performed using the non-destructive readout scheme
with 16 readouts that reduced the readout noise by a factor of
$1/\sqrt{16}$. The resulting readout noise was estimated to be
2.03 ADU. As the field of view (FOV) of \object{UM673} does not
contain nearby bright stars (within 1\arcmin~of \object{UM673}),
the brighter component A ($m_{\mathrm{R}}\sim$16.5 mag) of the
lensed system was selected as a guide star. The night of the
observations was very clear with natural seeing of about
0.5\arcsec~in the K band. \object{UM673} was observed at five
dither positions shifted by 9\arcsec. Every dither sequence was
repeated five times. The total number of images taken was 25 (22
of them were used in the following analysis). The exposure time of
each image was 150~s. Seeing of the images varied from
0.28\arcsec~to 0.46\arcsec. The airmass of \object{UM673} varied
from 1.35 to 1.88. We also observed nearby standard star Feige 16
($m_{\mathrm{H}}$=12.336~mag) from the MKO JHK Catalogue (Leggett
et al.~\cite{leggett2006}) with and without the AO correction. In
both cases the star was observed in one five-points dither
sequence. The exposure time of the observations without the AO
correction was 20~s, while the exposure time with the AO
correction was 5~s. The airmass and full width at half maximum
(FWHM) of the star in the observations with the correction were
1.85 and 0.2\arcsec, while without the correction they were 1.18
and 0.6\arcsec.

The near-infrared (NIR) observations of \object{UM673} with the
1.3m SMARTS telescope (CTIO, Chile) were carried out for a total
of 23 nights between August 16, 2009 and January 18, 2010 in the
J, H, and K bands. These data were acquired as a part of the ToO
observations carried out by National Central University, Taiwan.
The 1.3m SMARTS telescope is equipped with the dual-channel
optical/NIR CCD camera ANDICAM which has a NIR FOV of about
$2.2\arcmin\times2.2\arcmin$ and a pixel scale of 0.297\arcsec
$\rm{pixel^{-1}}$. On each observational night images were taken
in a series of four frames in the J, H, and K bands,
simultaneously with the optical data in the V, R, and I bands (see
Koptelova et al.~\cite{koptelova2012}). The exposure time of the
images in the J, H, and K bands varied from 132 to 225~s.

\section{Photometry of the AO Subaru data}\label{subaru}

The Subaru H-band image frames of \object{UM673} were reduced with
the IRAF\footnote{IRAF is distributed by the National Optical
Astronomy Observatories, which are operated by the Association of
Universities for Research in Astronomy, Inc., under cooperative
agreement with the National Science Foundation} scripts written
for the reduction of the IRCS data by Y.~Minowa\footnote{The
scripts can be found at the Subaru Data Reduction webpage
http://www.naoj.org/Observing/DataReduction/index.html}~(\cite{minowa2010})
(see also Minowa et al.~\cite{minowa2005}). The bad pixels were
removed by interpolation from adjacent good pixels using file
cam\_badpix.coo\footnote{This file is available at the Subaru IRCS
webpage http://www.naoj.org/Observing/Instruments/IRCS/} with the
positions of the IRCS bad pixels. We corrected the images for the
relative pixel offset and co-added them into a stack image using
median-filter coaddition.

\begin{figure}
\resizebox{\hsize}{!}{\includegraphics{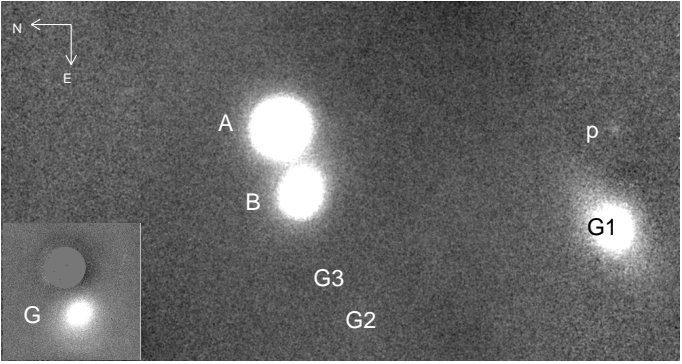}}
\caption{$17\arcsec \times 10\arcsec$ Subaru H-band stack image of
\object{UM673}. Lensing galaxy G after subtracting the light from
quasar images A and B is shown in a lower zoom window.}
\label{fig1}
\end{figure}

\begin{figure}
\resizebox{\hsize}{!}{\includegraphics{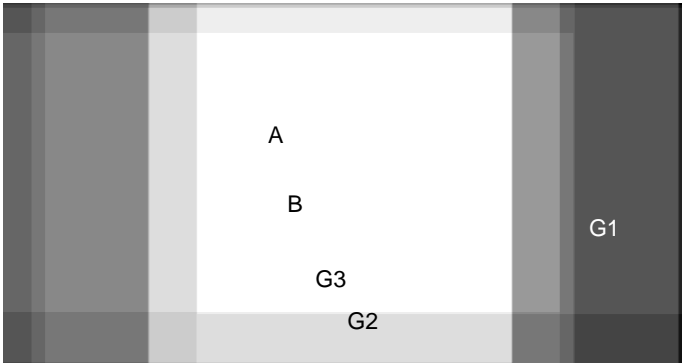}}
\caption{Exposure map of the \object{UM673} field presented in
Fig.~\ref{fig1}. The grey scale corresponds to the total
integration time.} \label{fig2}
\end{figure}

The $17\arcsec \times 10\arcsec$ subregion of the stack image is
shown in Fig.~\ref{fig1}. The FWHM of the stack image estimated
from component A of \object{UM673} is about 0.43\arcsec. A lower
left-corner window displays the light from the lensing galaxy
after subtracting quasar images A and B (see Sec.~\ref{subaru} for
details). Fig.~\ref{fig2} shows the exposure map of the stack
image presented in Fig.~\ref{fig1}. The white central region of
the map with the marked locations of quasar images A and B
corresponds to a maximum exposure time of 55 minutes.

To calibrate the fluxes, we measured a zero-magnitude using the
combined (stack) images of the standard star observed with the AO
correction, as in this case the airmass of the star and
\object{UM673} was almost the same. The H-band zero-magnitude was
determined to be $24.294\pm0.013$~mag for 1~ADU~ per second. We
also estimated a limiting magnitude for the point source detection
in the combined image of \object{UM673} using component A of the
quasar. The $5\sigma$ limiting magnitude measured as in
Minowa~(\cite{minowa2010}) was estimated to be about 22.5~mag at
an aperture radius of 0.3\arcsec. The $5\sigma$ surface brightness
limiting magnitude for the region of the combined image with the
longest total exposure was found to be about 21.6~mag
$\rm{arcsec^{-2}}$. The $5\sigma$ surface brightness limiting
magnitude with bright field galaxy G1 was estimated to be about
20.5 mag~$\rm{arcsec^{-2}}$.

To detect possible faint objects in the field of \object{UM673} we
ran SExtractor (Bertin \& Arnouts~\cite{bertin1996}) on the stack
image. The SExtractor search parameters were fixed to those used
by Minowa et al. (\cite{minowa2005}) for the faint-source
detection in the deep NIR observations with the Subaru telescope.
The SExtractor did not detect any faint objects except for a
low-level signal at position p (see Fig.~\ref{fig1}). This signal
results from the persistent after-image of quasar component A and
at first was misclassified by us as a new faint galaxy. The image
persistence of component A was noticed in every fourth image of
the five-image dither sequence. In other dither images the effect
was weaker or not present at all. Fig.~\ref{fig1} also shows the
locations of the two faint galaxies G2 and G3 detected in the HST
observations of \object{UM673} (Leh$\rm \acute{a}$r et
al.~\cite{lehar2000}). The H-band magnitudes of these galaxies as
measured by Leh$\rm \acute{a}$r et al.~(\cite{lehar2000}) are
about 20.65 and 21.78~mag. Their locations correspond to the
longest integration time in the Subaru stack image (see
Fig.~\ref{fig2}). However, we detected no objects at the HST
positions of G2 and G3 as they appear to be fainter than the
limiting magnitude achieved in the Subaru observations.

\begin{figure}
\resizebox{\hsize}{!}{\includegraphics{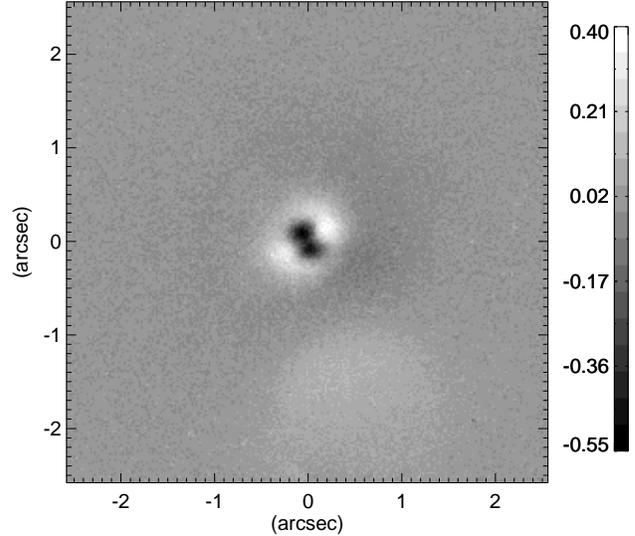}}
\caption{Residuals after subtracting the PSF model constructed
using observations of the standard star.} \label{fig3}
\end{figure}

The photometric analysis of the lensed system was performed with
the point spread function (PSF) fitting method. The model of the
PSF was constructed using component A of \object{UM673}. To
analyse the brightness distribution of the PSF, we extracted a
5\arcsec~subframe from the \object{UM673} frame centred on
component A. The light from component B in the subframe was
blocked using a circular mask. The PSF model was constructed in
two steps. First, the PSF was modelled as a combination of the
elliptical Gaussian and Moffat (Moffat~\cite{moffat1969})
profiles. In these calculations we used the brightness
distribution of the standard star to include all hard-to-model
features of the AO PSF into the PSF model. We blurred the star PSF
with the combined Gaussian-Moffat profile, so that the resulting
PSF was fitted to the brightness distribution of component A. The
residuals between the data and the model are shown in
Fig.~\ref{fig3}. The vertical bar demonstrates pixel counts in ADU
sec$^{-1}$ that can be compared with a sky background sigma of
0.018 ADU sec$^{-1}$. As can be seen, this PSF model cannot
describe the shape of the observed PSF in its centre. Therefore,
second, we replaced the central region (within a radius of
0.7\arcsec) of the best-fit Gaussian-Moffat PSF with the observed
brightness distribution of component A. This hybrid
analytical-and-empirical PSF was used for the photometric analysis
of the \object{UM673} lensed system.

\begin{table}
\caption{Relative astrometry of component B, the lens, and the
nearby bright galaxy.} \label{table1} \centering
\begin{tabular}{l c c c c}
\hline   ID  &  z      &  R.A.(\arcsec)    & DEC.(\arcsec) & H
(mag)\\\hline
A  &  2.72 & 0               &  0  &  $15.174\pm0.015$ \\
B  &  2.72 & $2.101\pm0.004$ & $-0.585\pm0.005$ &  $17.335\pm0.018$  \\
G  &  0.49 & $1.751\pm0.015$ & $-0.540\pm0.005$ &  $17.816\pm0.022$\\
G1 &  0.17 & $3.028\pm0.004$ & $-9.929\pm0.004$ &  $17.854\pm0.015$ \\
\hline
\end{tabular}
\end{table}
\begin{table}
\caption{Galaxy parameters: S\'{e}rsic index $n$; effective radius
$r_{\mathrm{e}}$; axis ratio $q$; and position angle $PA$.}
\label{table2} \centering
\begin{tabular}{l c c c c }
\hline   ID  &   $n$    & $r_{\mathrm{e}}(\arcsec)$ &
$q=\frac{b}{a} $& $PA$($^{\circ}$)\\\hline
G  &   $3.93\pm0.06$ & $0.86\pm0.08$ & $0.80\pm0.02$ & $30.5\pm3.4$\\
G1 &   $4.28\pm0.39$ & $0.85\pm0.09$ & $0.68\pm0.03$ &
$125.9\pm2.2$  \\\hline
\end{tabular}
\end{table}

The brightness profile of each of the quasar components was
modelled as the Dirac delta function convolved with the
analytical-and-empirical PSF. We approximated the delta function
using a narrow Gaussian with a sigma of 0.5 pixels (0.01\arcsec).
The model parameters describing the quasar components are the
fluxes of components A and B scaled with the PSF and their $x$ and
$y$ coordinates on the frame.

The galaxy brightness distribution was modelled with the S$\rm
\acute{e}$rsic profile of the form
\begin{equation}
I(r)=I_{\mathrm{e}}\exp\left\{-b_{\mathrm{n}}
\left[\left(\frac{r}{r_{\mathrm{
e}}}\right)^{\frac{1}{n}}-1\right]\right\} \label{sersic},
\end{equation}
where $I_{\mathrm{e}}$ is the intensity at effective radius
$r_{\mathrm{e}}$ (S$\rm \acute{e}$rsic~\cite{sersic1968}); $r^{2}
= x^{2} + y^{2}/q^{2}$ is specified by parameter $q$, the axis
ratio of the ellipse. Quantity $b_{\mathrm{n}}$ is a function of
shape parameter $n$ and defined so that $r_{\mathrm{ e}}$ encloses
half of the total light (see Ciotti~\cite{ciotti1991}). In the
calculations we express $I_{\mathrm{e}}$ in Eq.~\ref{sersic}
through the total luminosity $L_{\mathrm{tot}}$ (see also equation
2 in Ciotti~\cite{ciotti1991} and equation 2 in Graham \&
Driver~\cite{graham2005}). Thus, the S\'{e}rsic profile is
described with the following free parameters: centre coordinates
of the galaxy $x_{\mathrm{c}}$ and $y_{\mathrm{c}}$,
$L_{\mathrm{tot}}$, $r_{\mathrm{e}}$, $n$, $q$, and position angle
$PA$ of the galaxy major axis. The S\'{e}rsic model of the galaxy
was then convolved with the PSF.

\begin{figure}
\resizebox{\hsize}{!}{\includegraphics{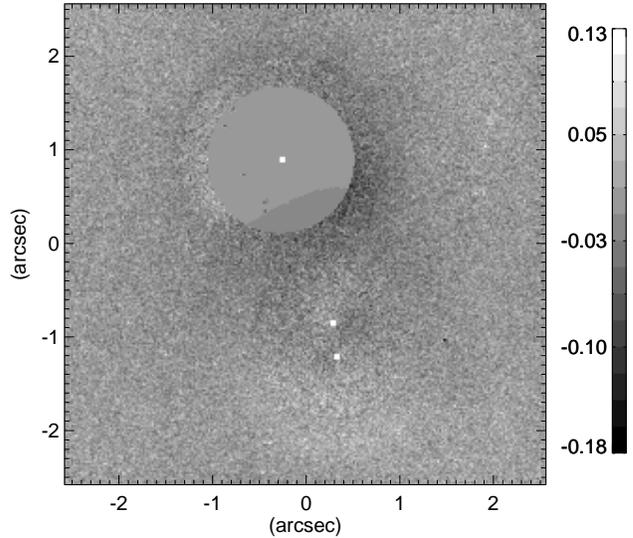}}
\caption{Residuals after subtracting the light of quasar images A
and B, and the lensing galaxy. The positions of A, B, and the
galaxy centre as calculated from the model are shown by white
dots.} \label{fig5}
\end{figure}

The solution for the best-fit photometric model of UM673
($\chi^{2}_{\mathrm{\nu}}\sim0.003$) is summarised in
Tables~\ref{table1} and \ref{table2}. (Note that we used minimum
change in $\chi^{2}_{\mathrm{\nu}}$ as the stopping criterion for
the minimisation procedure.) Table~\ref{table1} presents the
positions of component B and the lensing galaxy relative to
component A, and the calibrated magnitudes of the quasar
components and lensing galaxy. The tables also present the results
of the fit of galaxy G1 with the S\'{e}rsic profile and its
S\'{e}rsic magnitude. The magnitude errors in Table~\ref{table1}
are the errors of the fit, which also include the error of the
zero-magnitude. Fig.~\ref{fig5} shows residuals between the
observational data and the model. In this figure a circular region
at the location of component A corresponds to the central part of
the hybrid PSF described empirically. The lensing galaxy after
subtracting the light from quasar components A and B is shown in
the lower left corner of Fig.~1.

With the goal to identify any faint object at small impact
parameters from the quasar, we also considered another approach.
We subtracted the light from the quasar using the principal
component (PC) representation of the AO complex PSF as described
in Chun et al.~(\cite{chun2006, chun2010}). The PC decomposition
was applied to a sequence of the single-exposure subframes. We
analysed a sequence of 22 subframes of \object{UM673} that
contained both components A and B. In each subframe, component A
represents the best model for the PSF. The PCs were calculated by
diagonalising the covariance matrix between the 22 single-exposure
PSFs. The resulting PSF $P(x,y)$ in the combined image is
expressed as a linear combination of the PC components,
$P(x,y)=\sum_{i=1}^{22}a_{i}PC_{i}(x,y)$, where $a_{i}$ are the
coefficients that represent the importance of each of the PC modes
of the AO PSF in the combined image and quantify the variations of
the AO PSF with time. Then, the model, with $a_{i}$ as parameters,
was fitted to the observational data.

\begin{figure}
\resizebox{\hsize}{!}{\includegraphics{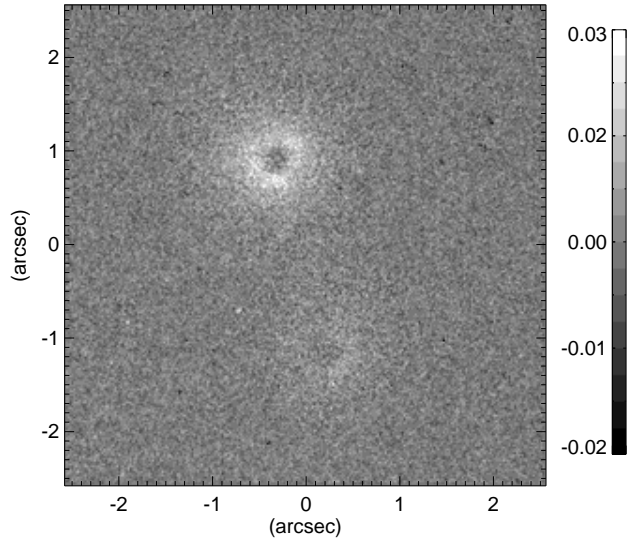}}
\caption{Residuals after subtracting the scaled PC modes from the
stack image.} \label{fig4}
\end{figure}

The residuals between the observational data and the PC model are
shown in Fig.~\ref{fig4}.
As can be seen in the figure, the PSF constructed using the PC
decomposition is able to remove most of the light from the quasar,
especially in the central region. The overall scatter in the
residuals appears to be small with higher values at the location
of component A and lower values at the location of component B. In
this residual map we do not detect any statistically significant
signal (above 1$\sigma$ detection) that might be produced by a
field object. However, we caution that some of the individual
images have irregular sky background that contributes to the sky
background of the stack image. We found that these irregularities
in the sky (especially small-scale irregularities) can lead to
mildly significant detections (below 1$\sigma$) in the residual
map. As most of these detections usually disappear after more
careful background correction, we considered them false
detections. If there were a real object with low significance, it
would be hard to distinguish its signal from these background
variations.

\section{Photometry of the CTIO data}\label{ctio_phot}

\begin{table*}
\caption{Positions and NIR magnitudes of the calibration stars.}
\label{table3} \centering
\begin{tabular}{l c c c c c}
\hline   ID  &   R.A.   & DEC. & J(mag) & H(mag) & K(mag)
\\\hline
$\alpha$  &   $1^{\mathrm{h}}45^{\mathrm{m}}16.55^{\mathrm{s}}$ &  $-9\degr46\arcmin29.54\arcsec$  & $16.823\pm0.159$ & $16.280\pm0.225$ & $16.106$\\
$\beta$   &   $1^{\mathrm{h}}45^{\mathrm{m}}15.83^{\mathrm{s}} $ &
$-9\degr45\arcmin43.44\arcsec$ & $17.216\pm0.015$ &
$16.683\pm0.054$ & $16.165\pm0.075$
\\\hline
\end{tabular}
\end{table*}
The CTIO NIR data of \object{UM673} were reduced in a similar way
as the Subaru data. We created sky flat-fields in the J, H, and K
bands using target frames taken at slightly offset positions. The
target frames were dark subtracted and divided by the normalised
sky flat-fields. The reduced frames taken on the same nights were
aligned and co-added.

The field of view of the CTIO NIR images contains only one star,
about 30\arcsec~south-west of \object{UM673}. We denote this star
as $\beta$. We used star $\beta$ to construct the PSF and to
calibrate fluxes. The fluxes of the quasar components relative to
the star were measured following the procedure described in
Koptelova et al.~(\cite{koptelova2008, koptelova2010,
koptelova2012}). Star $\beta$ itself does not have measured
calibrated magnitudes in the J, H, and K bands. To calibrate the
flux of star $\beta$ we used the available archive NIR
observations of \object{UM673} obtained with the VLT in July
2000\footnote{PI/CoI J. Hjorth et al. Proposal No. 65.O-0666(D).}.
The FOV of the VLT images is larger ($2.5\arcmin\times2.5\arcmin$)
and includes more stars than the CTIO images. One of the bright
stars in the VLT images of \object{UM673} has the measured 2MASS
J, H and K-band magnitudes (Skrutskie et
al.~\cite{skrutskie2006}). The position of this star (denoted as
$\alpha$) and its 2MASS magnitudes are given in
Table~\ref{table3}. We calibrated the flux of $\beta$ relative to
the flux of $\alpha$ (see Table~\ref{table3}). Finally, we
calibrated the fluxes of components A and B of \object{UM673}
relative to star $\beta$.

The NIR brightness of the lensing galaxy located closer to
component B of \object{UM673} is similar to the brightness of B
(see Table~\ref{table1} in Sect.~\ref{subaru} and Leh$\rm
\acute{a}$r et al.~\cite{lehar2000}). Therefore, we expect a
significant contribution of the galaxy flux to the flux of B in
the CTIO NIR data. This contribution cannot be estimated based on
the CTIO images alone, as their spatial resolution does not allow
for separation of the fluxes of component B and the lensing
galaxy. We estimated the galaxy flux contamination of B in the
CTIO measurements using the VLT NIR observations of
\object{UM673}$^{4}$, which are deeper and have higher angular
resolution (0.147\arcsec pixel$^{-1}$). From the VLT data we
measured the fluxes of A and B corrected for the galaxy flux. In
these measurements the galaxy brightness was modelled with the de
Vaucouleurs profile (de Vaucouleurs~\cite{deVaucouleurs1948}).
Then, the VLT data were rescaled and blurred, so as to match the
CTIO data in the spatial resolution and typical seeing. After
that, the fluxes of components A and B were measured again but
without the correction for the galaxy flux. From the comparison of
the fluxes of the quasar components measured in these two ways
(with and without the galaxy flux correction) we find that, within
errors of measurements, the galaxy does not contribute to the flux
of component A. But its contribution to the flux of component B is
significant and estimated to be 0.663, 0.836, and 0.999~mag in the
J, H, and K bands. These values were taken into account in the
CTIO photometry of component B.

The measured J- and H-band magnitudes of the quasar components are
given in Tables~\ref{table4} and \ref{table5}. We could not apply
the PSF photometry to the K-band data. The signal from
\object{UM673} and the PSF star in the CTIO K-band images was very
low as the K band suffered a very high background level.
Therefore, we instead measured the average K-band flux ratio
between the quasar components. In these measurements, the PSF was
constructed using component A of \object{UM673}. The K-band flux
ratio averaged over the observations taken between September 30
and October 28, 2009 is estimated to be $2.151\pm0.040$~mag. The
J- and H-band flux ratios averaged over the period of the
observations are $2.183\pm0.006$ and $2.162\pm0.010$.
\begin{table}
\caption{Photometry of components A and B of \object{UM673} in the
J band.} \label{table4} \centering
\begin{tabular}{c c c c}
\hline   JD  &   seeing(\arcsec)   & $\rm J_{\mathrm{A}}$(mag)  &
$\rm J_{\mathrm{B}}$(mag)
\\\hline
2455063.801  &   1.8 & $ 15.356\pm0.013$ & $17.535\pm0.021$ \\
2455064.825  &   1.3 & $ 15.362\pm0.005$ & $17.541\pm0.006$ \\
2455069.816  &   1.4 & $ 15.407\pm0.016$ & $17.555\pm0.023$ \\
2455070.769  &   1.4 & $ 15.382\pm0.023$ & $17.569\pm0.029$ \\
2455116.652  &   1.3 & $ 15.395\pm0.006$ & $17.604\pm0.013$
\\\hline
\end{tabular}
\end{table}
\begin{table}
\caption{Photometry of components A and B of \object{UM673} in the
H band.} \label{table5} \centering
\begin{tabular}{c c c c}
\hline   JD  &   seeing(\arcsec)   & $\rm H_{\mathrm{A}}$(mag)  &
$\rm H_{\mathrm{B}}$(mag)
\\\hline
2455070.792  &   1.4 & $ 15.210\pm0.012$ & $17.372\pm0.016$ \\
2455082.758  &   1.4 & $ 15.229\pm0.006$ & $17.380\pm0.028$ \\
2455083.796  &   1.2 & $ 15.218\pm0.013$ & $17.375\pm0.015$ \\
2455090.789  &   1.3 & $ 15.182\pm0.018$ & $17.359\pm0.027$ \\
2455095.750  &   1.5 & $ 15.189\pm0.014$ & $17.370\pm0.024$ \\
2455100.658  &   1.3 & $ 15.181\pm0.016$ & $17.376\pm0.038$ \\
2455105.730  &   1.6 & $ 15.202\pm0.009$ & $17.377\pm0.029$ \\
2455116.675  &   1.3 & $ 15.186\pm0.012$ & $17.342\pm0.021$
\\\hline
\end{tabular}
\end{table}

The employed procedure of measuring the uncontaminated flux of B
based on the VLT data is simple and accurate at the same time.
However, we note that there are other approaches as well. Ricci et
al.(\cite{ricci2013}) adopted an alternative way of estimating the
galaxy flux contribution to B in their photometry of the optical
\object{UM673} data obtained with the Danish 1.54m telescope. In
the method the galaxy flux is constrained using the HST positions
and photometry.

\section{Intrinsic flux ratio between A and B}\label{extinction}

The observed flux ratio between the quasar components might differ
from the flux ratio produced by lensing on the galaxy. We refer to
the flux ratio resulting from lensing as the intrinsic flux ratio.
Given that the fluxes of the quasar components are corrected for
the time delay, the intrinsic flux ratio can be altered by
microlensing on the stars in the lens, by dust extinction in the
lensing galaxy or other objects between the quasar and observer.
We assume that the observed fluxes of A and B differ from their
intrinsic fluxes mainly because of the dust extinction, as there
was no clear microlensing detected in the previous observations of
\object{UM673} (see Wisotzki et al.~\cite{wisotzki2004}; Koptelova
et al.~\cite{koptelova2008, koptelova2010, koptelova2012}). We
also explain the microlensing-like behaviour (the
bluer-when-brighter) of component A found by Nakos et
al.~(\cite{nakos2005}) by the quasar intrinsic variability and not
by microlensing (see also the discussion in Koptelova et
al.~\cite{koptelova2012}).

The amount of dust reddening at a given wavelength can be
estimated from the observed multi-band flux ratios of the quasar
components by adopting a certain extinction law in the lensing
galaxy. For this purpose we used the V, R, I, J, H and K-band flux
ratios between components A and B of \object{UM673}. The optical
flux ratios are taken from the measurements of Koptelova et
al.~(\cite{koptelova2012}), while the NIR flux ratios are those
from the CTIO observations presented in Sect.~\ref{ctio_phot}. The
flux ratios in the optical bands were averaged over several epochs
to eliminate the effect of the time delay on the single-epoch
measurements. As shown in Koptelova et al.~(\cite{koptelova2012}),
the multi-epoch average flux ratio between components A and B is a
good approximation for their time-delay-corrected flux ratio. The
NIR light curves are too short to be corrected for the time delay.
In the calculation of the NIR flux ratios, we assumed that the
effect of the time delay is weak and within the photometric
errors. This is a valid assumption as quasar brightness variations
in the NIR bands are smaller than in the optical bands, and UM673
itself was very quiet during our NIR observations.

We also used the B-band flux ratio measured based on different
single-epoch observations of \object{UM673}. The average B-band
flux ratio was calculated based on our own observations of
\object{UM673} with the 1.3m CTIO telescope and the measurements
published earlier by El\'{i}asd\'{o}ttir et
al.~(\cite{eliasdottir2006}) and Mosquera et
al.~(\cite{mosquera2011}). The CTIO B-band flux ratio between A
and B was estimated using the observations obtained on August 26,
September 30, and October 11, 2009. It was measured to be
$2.246\pm0.008$~mag. The multi-epoch weighted average flux ratio
adopted by us for the calculations is $2.320\pm0.025$~mag.

The J, H and K-band flux ratios were corrected for the galaxy
contribution as described in Sect.~\ref{ctio_phot} of this paper.
The galaxy contribution to the optical fluxes is difficult to
estimate based on the low-resolution ground-based observations
alone. Here, we estimated the galaxy contribution in the optical
bands using the following strategy: it is known from observations
of \object{UM673} that the galaxy flux contribution to component B
increases from the B band, in which it is negligible (we can
likely assume it to be zero), to the K band, in which it becomes
much more significant (about 1~mag). The contamination of
component B by the galaxy light was also noticed in the
spectroscopic observations of Surdej et al.~(\cite{surdej1988}),
who found an excess in the continuum radiation of the component at
$\lambda>5800\AA$ (this lower wavelength roughly corresponds to
the V band) due to the galaxy light contribution. Therefore, the
galaxy contamination of component B, if not taken into account,
introduces a certain trend into the flux ratio-wavelength
dependance. We estimated this trend using the B, J, H and K-band
measurements of the galaxy flux contribution to component B. The
dependence of the contribution on wavelength was fitted with a
second-order polynomial. From the resulting curve we estimated the
amount of the galaxy flux contribution to each band between the B
and K bands. Then the optical flux ratios were corrected for this
amount by subtracting the second-order polynomial from the
observed flux ratios, measured without the galaxy correction.

\begin{figure}
\resizebox{\hsize}{!}{\includegraphics{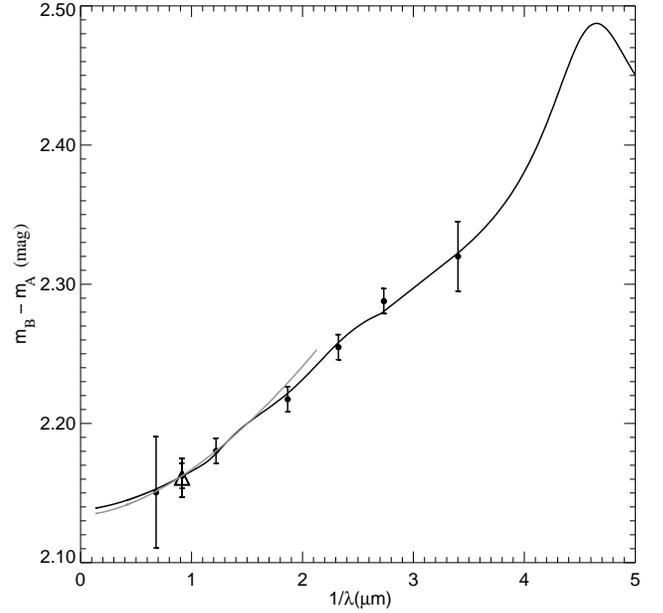}}
\caption{Magnitude difference between components A and B of
\object{UM673} as a function of the rest-frame wavelength. The fit
to the NIR measurements of the magnitude differences is shown by a
grey line, while the fit to the optical-NIR measurements is shown
by a black line. } \label{fig6}
\end{figure}

From the analysis we find that the galaxy contribution to the flux
of component B is about 0.098, 0.204, and 0.363~mag in the V, R,
and I bands. The corrected flux ratios at the optical wavelengths
and the measured NIR flux ratios are shown in Fig.~\ref{fig6} by
circles. The H-band flux ratio measured from the Subaru
observations of \object{UM673} is marked by a triangle. As can be
seen, the Subaru flux ratio measured based on the single-epoch
observations agrees well with the CTIO H-band flux ratio measured
based on the multi-epoch data. The resulting flux ratios in the B,
V, R, I, J, H, and K bands were used to measure the differential
dust extinction between the lines of sights to components A and B
and also to estimate their intrinsic flux ratio.

We analysed the extinction properties of the lensing galaxy using
the NIR extinction law (see review of Mathis~\cite{mathis1990} and
references therein) and the parametrisation of the Galactic
extinction law given in Cardelli et al.~(\cite{cardelli1989}). The
magnitude difference $\Delta m^{\mathrm{BA}}(\lambda)$ and the
flux ratio between the quasar components are related as $\Delta
m^{\mathrm{BA}} = 2.5\ast\rm lg(\it
f_{\mathrm{A}}/f_{\mathrm{B}})$. The intrinsic magnitude
difference between the quasar components is defined as $\Delta
m_{0}^{\mathrm{BA}}=m_{\mathrm{0}}^{\mathrm{B}}(\lambda)-m_{\mathrm{0}}^{\mathrm{
A}}(\lambda)$ and does not depend on wavelength. The differential
extinction at wavelength $\lambda$ is determined by the
discrepancy between the observed and intrinsic magnitude
differences as $\Delta A(\lambda) = \Delta
m^{\mathrm{BA}}(\lambda)-\Delta m^{\mathrm{BA}}_{\mathrm{0}}$.
Generally, this differential extinction accounts for all possible
causes of reddening of the quasar components, dust reddening in
the lensing galaxy, and in the intervening objects along the line
of sight to each of the components.

The NIR magnitude difference between components A and B is
represented as follows:
\begin{equation}
      \Delta m^{\mathrm{BA}}(\lambda) = \Delta m_{\mathrm{0}}^{\mathrm{BA}}+\Delta A(J)\left(\frac{\lambda}{1.25\mu \rm m}\right)^{-1.7},
      \label{magdiffnir}
   \end{equation}
with $\Delta A(J) = A^{\mathrm{B}}(J)-A^{\mathrm{A}}(J)$ (see
Mathis~\cite{mathis1990}). The model defined by
Eq.~\ref{magdiffnir} has two parameters, $\Delta
m_{\mathrm{0}}^{\mathrm{BA}}$ and $\Delta A(J)$. From the fit, we
calculated $\Delta m_{\mathrm{0}}^{\mathrm{BA}}=2.134\pm0.024$~mag
and $\Delta A(J)=0.023\pm0.014$~mag
($\chi^{2}_{\mathrm{\nu}}\sim0.0$). The differential extinction in
the H and K bands was estimated to be about 0.015 and 0.009~mag.

For the extinction curve parameterised as in Cardelli et
al.~(\cite{cardelli1989}), the magnitude difference between
components A and B is determined by the intrinsic magnitude
difference $\Delta m_{\mathrm{0}}^{\mathrm{BA}}$, the differential
extinction $\Delta A(V)$, and the differential colour excess
$\Delta E(B-V)$ as follows:
\begin{equation}
      \Delta m^{\mathrm{BA}}(\lambda) = \Delta m_{\mathrm{0}}^{\mathrm{BA}}+\Delta A(V)a(\lambda^{-1})+\Delta
      E(B-V)b(\lambda^{-1}),
      \label{magdiff}
   \end{equation}
where $\Delta A(V) = A^{\mathrm{B}}(V)-A^{\mathrm{A}}(V)$ and
$\Delta E(B-V) = E^{\mathrm{B}}(B-V)-E^{\mathrm{A}}(B-V)$. From
the fit between the observed and model magnitude differences, we
calculated $\Delta
m_{\mathrm{0}}^{\mathrm{BA}}=2.138\pm0.015$~mag, $\Delta
A(V)=0.081\pm0.018$~mag, and $\Delta E(B-V)=0.035\pm0.003$~mag
($\chi^{2}_{\mathrm{\nu}}\sim0.3$).

Thus, the values of $\Delta m_{\mathrm{0}}^{\mathrm{BA}}$ derived
from the NIR and Galactic extinction curves agree well. The
effective ratio of the total-to-selective extinction from the data
is $\Delta A(V)/\Delta E(B-V)\simeq2.29$. This value is in the
range of the $R_{\mathrm{V}}$ values measured by Patil et
al.~(\cite{patil2007}) for a sample of 26 early-type galaxies.

\section{\object{UM673} lens model}\label{model}

In this section we explore lens models for the \object{UM673}
system. We analysed the models to identify the source of shear and
to estimate the probable location of the nearby perturber. The
models were constrained using the positions of the two quasar
components, their flux ratio, and a time delay of $89\pm11$~days
measured in Koptelova et al.~(\cite{koptelova2012}). In the
calculations we used positions of the quasar components and
properties of the lensing galaxy as measured from the Subaru
images of \object{UM673} (see Tables~\ref{table1} and
\ref{table2}).  The adopted value of $H_{0}$ is $74.2 \pm 3.6
$~km~s$^{-1}$Mpc$^{-1}$ as measured in Riess et
al.~(\cite{riess2009})

The lensing galaxy was modelled as a singular isothermal ellipsoid
(SIE) (see Kassiola \& Kovner \cite{kassiola1993}; Kormann et
al.~\cite{kormann1994}; Keeton \& Kochanek~\cite{keeton1998}). The
properties of the SIE model and corresponding gravitational
potential are given in Keeton \& Kochanek~(\cite{keeton1998}) and
Keeton~(\cite{keeton2001}). The model parameters were calculated
by minimising the following function:
\begin{equation}
\chi^{2}=\chi^{2}_{\mathrm{posn}}+\chi^{2}_{\mathrm{flux}}+\chi^{2}_{\mathrm{delay}},
  \label{totalchi}
\end{equation}
where $\chi^{2}_{\mathrm{posn}}$ is the $\chi^{2}$ term for the
image positions; $\chi^{2}_{\mathrm{flux}}$ is the $\chi^{2}$ term
for the image flux ratio, and $\chi^{2}_{\mathrm{delay}}$ is the
$\chi^{2}$ term for the time delay. In our calculations the
$\chi^{2}_{\mathrm{posn}}$ term was evaluated in the source plane
as described in Keeton~(\cite{keeton2010}). The
$\chi^{2}_{\mathrm{flux}}$ term is defined as
$\chi^{2}_{\mathrm{flux}}=(f_{\mathrm{B}}/f_{\mathrm{A}}-\mu_{\mathrm{B}}/\mu_{\mathrm{A}})^{2}/\sigma_{\mathrm{flux}}^{2}$,
where $f_{\mathrm{B}}/f_{\mathrm{A}}$ and
$\mu_{\mathrm{B}}/\mu_{\mathrm{A}}$ are the observed and model
flux ratio. The $\chi^{2}_{\mathrm{delay}}$ term was calculated as
$\chi^{2}_{\mathrm{delay}}=(\Delta
t_{\mathrm{BA}}^{\mathrm{obs}}-\Delta
t_{\mathrm{BA}}^{\mathrm{mod}})^{2}/\sigma_{\mathrm{delay}}^{2}$.
The models were fitted using our routines written in the
interactive data language (IDL).
\begin{table}
\caption{Model 1 results: the SIE model with an external shear.} 
\label{model_solutions1} 
\centering 
\begin{tabular}{c c c} 
\hline\hline 
Parameter & SIE+$\gamma$ & SIE+$\gamma$  \\ 
\hline 
$b(\arcsec)$            & $1.155\pm0.009$ & $1.157\pm0.009$ \\ 
$q$                     & $0.792\pm0.004$ & $0.784\pm0.003$ \\
$PA(\degr)$           & $56.7\pm0.5$ & $57.1\pm0.3$ \\
$\gamma$                & $0.075\pm0.003$ & $0.068\pm0.003$ \\
$\theta_{\mathrm{\gamma}}(\degr)$ & $53\pm2$ & $51\pm2$ \\
\hline 
$\mu_{\mathrm{B}}/\mu_{\mathrm{A}}$ &  $0.140$ & $0.141$  \\
$\Delta t_{\mathrm{BA}}$(days)   &  $115$   & $114$ \\
\hline 
$\chi^{2}/N_{\mathrm{DOF}}$      &  0.0/-2 & 5.3/-1 \\
\hline 
\end{tabular}
\end{table}
\begin{table}
\caption{Model 2 and 3 results: the SIE-and-SIS model with an external shear.} 
\label{model_solutions2} 
\centering 
\begin{tabular}{c c c} 
\hline\hline 
Parameter & SIE+SIS+$\gamma$(2) & SIE+SIS+$\gamma$(3)  \\ 
\hline 
$b(\arcsec)$            & $1.091\pm0.014$ & $ 0.910^{+0.036}_{-0.011}$ \\ 
$b_{\mathrm{p}}(\arcsec)$        & $0.601\pm0.010$ & $0.648^{+0.146}_{-0.087}$ \\
$x_{\mathrm{p}}(\arcsec)$        & -- & $2.080^{+0.420}_{-0.280}$ \\
$y_{\mathrm{p}}(\arcsec)$        & -- & $0.745^{+0.478}_{-0.197}$ \\
$\gamma$                & $0.042\pm0.003$ & $0.173^{+0.024}_{-0.009}$ \\
$\theta_{\mathrm{\gamma}}(\degr)$ & $51\pm2$ & $87^{+1}_{-10}$ \\
\hline 
$\mu_{\mathrm{B}}/\mu_{\mathrm{A}}$ &  $0.141$ & $0.140$  \\
$\Delta t_{\mathrm{BA}}$(days)   &  $116$   & $90$ \\
\hline 
$\chi^{2}/N_{\mathrm{DOF}}$      &  6.0/0 & 0.1/-2 \\
\hline 
\end{tabular}
\end{table}

\begin{figure}
\resizebox{\hsize}{!}{\includegraphics{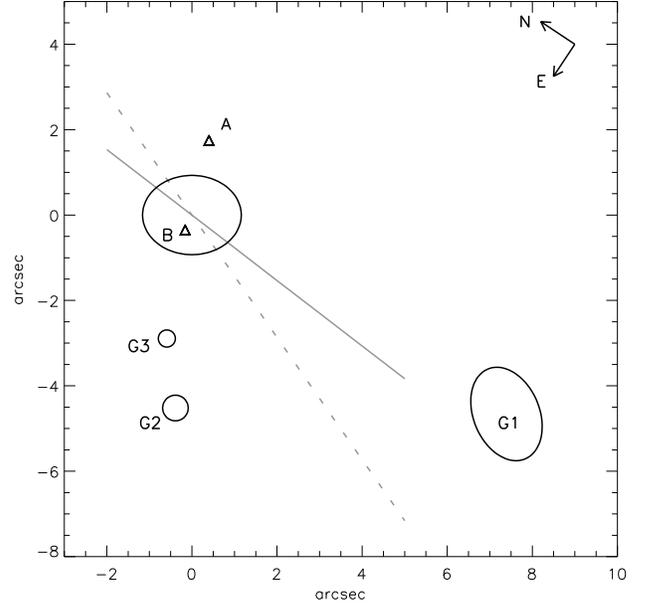}}
\caption{Locations of the nearby objects in the vicinity of UM673.
The galaxies are shown with their critical curves as if they were
isolated objects. The orientations of the total and the largest
shear in the field are shown by dotted and solid grey lines.}
\label{fig7}
\end{figure}

First, following Keeton et al.~(\cite{keeton1998}) and Leh$\rm
\acute{a}$r et al.~(\cite{lehar2000}), we considered the SIE lens
model with an external shear (SIE+$\gamma$, see Kochanek
\cite{kochanek1991}; Bernstein \& Fischer \cite{bernstein1999};
Keeton \cite{keeton2001}; Schneider et al.~\cite{saas33}). We
denote this model as Model 1. The introduction of the external
shear is justified by the presence of the nearby galaxies detected
in the deep HST and Subaru images of \object{UM673}. The locations
of the nearby objects are illustrated in the scheme presented in
Fig.~\ref{fig7}. In the figure, the major axis of the lensing
galaxy is aligned with the horizontal axis. The scheme shows the
orientation of the total external shear as measured in Leh$\rm
\acute{a}$r et al.~(\cite{lehar2000}) from all objects within
20\arcsec~of \object{UM673} (marked by a dashed grey line). The
direction to the largest external shear is shown by a solid grey
line (see Table~4 in Leh$\rm \acute{a}$r et al.~\cite{lehar2000}).
Model~1 has the following free parameters: mass of the lens $b$,
axis ratio $q$ and position angle $PA$ of the galaxy, shear
strength $\gamma$ and its orientation $\theta_{\mathrm{\gamma}}$,
and actual position of the quasar. The galaxy ellipticity and
position angle are chosen to be free parameters because models
with fixed $q$ and $PA$ do not fit well ($\chi^{2}>100$, see also
Leh$\rm \acute{a}$r et al.~\cite{lehar2000}).

\begin{figure*}
\centering
\includegraphics[width=15cm]{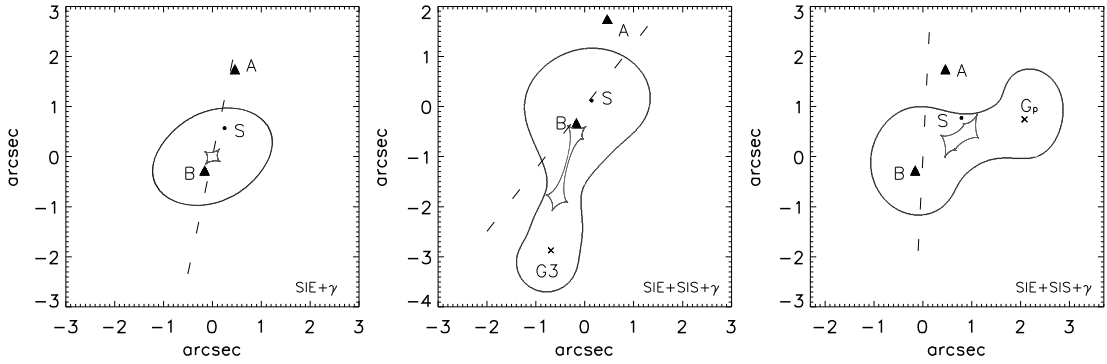}
\caption {Critical curves (black) and caustics (grey) for the lens
models presented in Tables~\ref{model_solutions1} and
\ref{model_solutions2} (right: Model 1, middle: Model 2, left:
Model 3). The positions of quasar images A and B are shown by
triangles. The source position is marked by a circle. Dashed lines
depict the orientation of the external shear. Crosses indicate the
positions of the perturbers.} \label{fig8}
\end{figure*}

Table~\ref{model_solutions1} summarises the results of the fit.
The solution for the Model~1 parameters calculated using
Eq.~\ref{totalchi} is given in the second column of the table. The
critical lines for this solution are shown in the left panel of
Fig.~\ref{fig8}. The values of the model parameters presented in
the first column of Table~\ref{model_solutions1} were calculated
for the SIE+$\gamma$ model constrained only by the image positions
and the flux ratio
($\chi^{2}=\chi^{2}_{\mathrm{posn}}+\chi^{2}_{\mathrm{flux}}$).
The errors on the model parameters were obtained from the
corresponding covariance matrix calculated at the last iteration
of the minimisation process.

As can be seen in Table~\ref{model_solutions1}, the position angle
of the lensing galaxy in Model~1 is different from the angle of
the observed galaxy light distribution (see Table~\ref{table2}).
The shear is oriented almost along the line that connects
components A and B. There are two faint galaxies, G2 and G3,
detected in the HST images of \object{UM673} along this direction
(see Fig.~\ref{fig7}). Therefore, from Model 1 we conclude that
the mass towards G2 and G3 probably influences the orientation and
ellipticity of the lensing potential in the model.

We also considered a lens model with galaxy parameters $q$ and
$PA$ fixed to the values measured from the observations. In this
model, we assumed that G3, which is closer to the main lens, has a
stronger effect on its potential. The mass concentration towards
G3 was modelled as a singular isothermal sphere (SIS) with one
free parameter, mass $b_{\mathrm{p}}$. The lensing effect from
other nearby galaxies at the position of the lens was described by
the external shear parameters. We denote this model as Model 2.
The solution for Model 2 is summarised in the first column of
Table~\ref{model_solutions2}. The critical lines for the model are
shown in the middle panel of Fig.~\ref{fig8}. In Model 2, the G3
perturber has a moderate mass and there is also a source of shear
at an angle of about 50\degr~measured counterclockwise from the
galaxy major axis.

In both Model 1 and Model 2, the mass at the locations of G2 and
G3 has no effect on the time delay. To examine which model of the
lens might lead to the observed delay of about 89 days, we
considered a model with a source of shear at the location of
galaxy G3 and an arbitrarily located SIS perturber of unknown mass
$b_{\mathrm{p}}$ (labelled $G_{\mathrm{p}}$). We denote this model
as Model~3. The model has seven parameters and therefore is
underconstrained. Two of the parameters, the shear strength and
its orientation, can be estimated. We estimated the shear strength
to be $\gamma=b_{\mathrm{G3}}/2r_{\mathrm{G3}}$, where
$b_{\mathrm{G3}}$ and $r_{\mathrm{G3}}$ are the mass and distance
of G3 from the main lens. The shear angle was assumed to be equal
to the observed angle towards G3 within an uncertainty of 10\degr.
From the fit we find that in this model the mass of the main lens
is smaller than in Model 1. The presence of the SIS perturber of a
moderate mass near the lens also makes the total mass between A
and B less concentrated. Model 3 predicts similar positions of the
quasar images and their flux ratio as the previous two models. At
the same time, its mass configuration favours shorter time delays.

\begin{figure*}
\centering
\includegraphics[width=15cm]{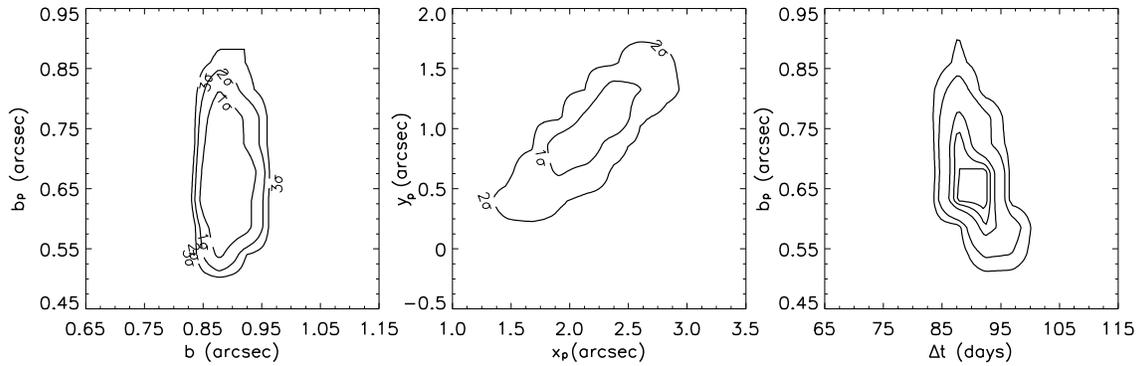}
\caption {Contours of the joint probability distribution in the
plane of lens mass $b$ and perturber mass $b_{\mathrm{p}}$ (left);
in the plane of the $x_{\mathrm{p}}$ and $y_{\mathrm{p}}$
perturber coordinates relative to the main lens (middle); in the
plane of time delay $\Delta t$ and perturber mass $b_{\mathrm{p}}$
(right). } \label{fig9}
\end{figure*}

To assess dependencies between the parameters in Model~3 and
estimate the parameter uncertainties we performed a statistical
analysis of the solutions. We minimised the $\chi^{2}$ function
for~500 realisations of Model 3. Each time we added Gaussian
distributed errors to the input shear strength and its
orientation. We also added the Gaussian error to the position of
the SIS perturber relative to the lensing galaxy. In this
analysis, the probable location of the SIS perturber was
constrained by only considering the solutions that favour time
delays shorter than 100 days. (In the configurations with longer
delays, perturber mass $b_{\mathrm{p}}$ becomes unimportant, and
the configurations closely resemble Models 1 and 2.)

Fig.~\ref{fig9} illustrates the result of the analysis. The two
first panels of the figure show contours corresponding to the
$1\sigma$, $2\sigma$, and $3\sigma$ confidence intervals in the
plane of two parameters: mass of the main lens $b$ and mass of the
perturber $b_{\mathrm{p}}$; coordinates $x_{\mathrm{p}}$ and
$y_{\mathrm{p}}$ of the perturber, respectively. (The $3\sigma$
contour in the middle panel is not shown because it is only poorly
resolved.) As can be seen in Fig.~\ref{fig9}, the mass of the main
lens in the parameter realisations of Model 3 is localised in a
narrow range of 0.9-1.0\arcsec. At the same time, the range of the
possible masses of the perturber is wider because $b_{\mathrm{p}}$
depends on the position of the perturber relative to the main
lens. The models with the larger distance between the lens and
perturber predict a higher mass of the perturber. As can be seen
in the right panel of Fig.~\ref{fig9}, which shows contours of the
equal probability distribution in the plane of the time delay and
perturber mass, the delay is shorter for the perturber with the
higher mass. But at the same time, the delay can also be shorter
in the models with the pertuber with the smaller mass located
closer to the main lens. The most probable values of the Model 3
parameters and their $1\sigma$ uncertainties are summarised in
Table~\ref{model_solutions2}. The corresponding critical lines for
Model 3 are shown in the right panel of Fig.~\ref{fig8}.

In Models 2 and 3, perturber $G_{\mathrm{p}}$ is assumed to be
located in the plane of the main lensing galaxy. To test the model
predictions for the case when $z_{\mathrm{G_{p}}}>z_{\mathrm{l}}$,
we also analysed the two-plane lens model reviewed in Schneider et
al.~(\cite{schneider1992}). In this analysis we set the SIS
perturber at different redshifts
$z_{\mathrm{G_{p}}}>z_{\mathrm{l}}$ and adopted the best solution
of Model 3 as an initial guess for the model parameters. As in
Model 3, the external shear was associated with galaxy G3. From
the calculations we find that the impact parameter of the
perturber from the quasar line of sight becomes smaller as its
redshift increases. These configurations produce a noticeable
lensing effect (see also the discussion in
Keeton~\cite{keeton2003} and Momcheva et al.~\cite{momcheva2006}).
For the case when $z_{\mathrm{G_{p}}}=1.63$ (the redshift of the
DLA absorber), we adopted one additional constraint assuming that
the position of quasar image A or, alternatively, quasar image B
in the perturber's plane is close to the position of
$G_{\mathrm{p}}$ within a conservative uncertainty of 0.1\arcsec.
The better solution ($\chi^{2}/N_{\mathrm{DOF}}\simeq$13/0)
corresponds to the configuration when $G_{\mathrm{p}}$ is closer
to image B in the perturber's plane. However, this model probably
forms the third image of the quasar because the predicted position
of $G_{\mathrm{p}}$ in this model is very close to the quasar line
of sight ($\sim0.15$\arcsec). The formation of the third image can
probably be suppressed if we assume a density profile shallower
than the isothermal model for the perturber (e.g., the NFW density
profile, Navarro et al.~\cite{navarro1996}). The other possibility
is that the perturber is located at a redshift lower than
$z_{\mathrm{G_{p}}}=1.63$ and, correspondingly, at the larger
impact parameter from the quasar line of sight.

\section{Summary and discussion} \label{discussion}

We presented new high-resolution AO observations of the
double-lensed quasar \object{UM673} obtained with the Subaru
telescope in the H band. Based on the Subaru data, we measured
optical properties of the lensing galaxy and the nearby bright
galaxy in the field of \object{UM673}. The measured H-band
magnitudes of the galaxies (17.816 and 17.854~mag, respectively)
agree well with the magnitudes previously estimated based on the
HST data (Leh$\rm \acute{a}$r et al.~\cite{lehar2000}).

We also presented results of the new near-infrared observations of
\object{UM673} obtained with the 1.3m SMARTS telescope at the CTIO
observatory in 2009. Based on these observational data, we
measured weighted J, H and K-band magnitude differences between
the \object{UM673} components of about $2.180\pm0.006$,
$2.162\pm0.009$, and $2.151\pm0.040$~mag. The flux ratios between
components A and B estimated from the analysis of the archive VLT
J, H and K-band images of \object{UM673} are $2.270\pm0.044$,
$2.216\pm0.057$, and $2.187\pm0.012$~mag.

From the comparison of the CTIO, VLT, HST NIR flux ratios and the
Gemini NIR flux ratio presented in Fadely \&
Keeton~(\cite{fadely2011}) we found a difference of up to 0.1~mag
between different measurements. The older observations (2000 and
earlier) with the VLT and especially with the HST result in higher
estimates of the NIR flux ratios than the more recent CTIO and
Subaru observations. This disagreement cannot be fully explained
by the differences in the photometric procedure. (In our test
analysis we measured the same HST H-band flux ratio as Leh$\rm
\acute{a}$r et al.~(\cite{lehar2000}).) As there were no detailed
NIR observations of \object{UM673} before 2000 to explain this
discrepancy, we connected this difference with the possible
evolution of the flux ratio with time (for example because of
quasar variability), although systematic errors in the estimation
of the flux ratio cannot be entirely excluded.

We analysed the dependence of the flux ratio on wavelength based
on recent extensive multi-band observations (see Koptelova et
al.~\cite{koptelova2008, koptelova2010, koptelova2012}; Ricci et
al.~\cite{ricci2013}; Oscoz et al.~\cite{oscoz2013}) and our new
NIR observations of \object{UM673}. From this analysis we measured
the extinction properties of the lensing galaxy and the intrinsic
flux ratio between the \object{UM673} components produced by
lensing on the galaxy. We estimated the effective ratio of the
total-to-selective extinction in the V band to be $\simeq2.29$.
This value agrees well with the much more precise measurement of
$R_{\mathrm{V}}$ in the early-type lensing galaxy of the system
\object{SBS 0909+532} (see Motta et al.~\cite{motta2002}). It also
agrees well with the results of Patil et al.~(\cite{patil2007}),
who studied extinction properties of early-type galaxies and
measured $R_{\mathrm{V}}$ to be in a range of 2.03-3.46 with an
average of 3.02. The intrinsic flux ratio between the
\object{UM673} components was estimated to be $2.138\pm0.015$~mag.
This value agrees better with the spectroscopic observations of
Wisotzki et al.~(\cite{wisotzki2004}) than the intrinsic flux
ratio previously estimated from the HST observations of
\object{UM673} in Falco et al.~(\cite{falco1999}).

The time delay between the two UM673 components as measured by
Koptelova et al.~(\cite{koptelova2012}) and Oscoz et
al.~(\cite{oscoz2013}) based on different datasets is somewhat
shorter than the delay predicted by the single-lens model. The
shorter delay leads to a value of the Hubble parameter that is
higher than the results of other independent measurements (e.g.,
Riess et al. \cite{riess2009}; Plank collaboration XVI
2013~\cite{ade2013}). This indicates a possible additional lensing
effect from nearby objects or objects on the line of sight to the
quasar (see, e.g., Keeton et al. \cite{keeton2000}; Keeton \&
Zabludoff \cite{keeton2004}). Leh$\rm \acute{a}$r et
al.~(\cite{lehar2000}) have measured a total convergence of
$k_{\mathrm{p}}$=0.138 from the nearby objects in the field of
\object{UM673} based on the previous HST observations. However,
this value cannot account for the shorter time delay and might be
underestimated. The total external convergence can probably be
measured with a better accuracy, e.g., by adopting the larger
radius around \object{UM673} for the analysis of the external
tidal effects or by more detailed imaging of the \object{UM673}
nearby field.

We performed new detailed imaging of the \object{UM673} assuming
that the additional mass is associated with a high-redshift
line-of-sight object at a small impact parameter from the quasar
(such as the host galaxy of the DLA absorber discovered by Cooke
et al.~\cite{cooke2010}). Imaging of the high-z DLA host galaxy
requires long integration times at the wavelength of the DLA
emission as well as special techniques that allow for removing the
quasar light contamination. In the current analysis of the 55-min
Subaru H-band image of the lensed quasar, we applied the principal
component representation of the complex AO PSF to subtract the
light from the \object{UM673} quasar components. We did not
convincingly detect any faint objects within 5\arcsec~of
\object{UM673} in the residual image.

Analysis of the imaging data was accompanied by the model analysis
of the lens. From the modelling we identified the source of shear
and estimated the probable location of nearby perturbing object
$G_\mathrm{p}$, the DLA host galaxy candidate. The shear is
probably associated with the field galaxies G2 and G3 seen in the
HST observations of \object{UM673}, while the observed time delay
can only be reproduced by the model with an additional perturbing
mass between components A and B (Model 3). The inferred total mass
of the lensing galaxy and the perturber is consistent with the
mass of the lens constrained by the positions of the
\object{UM673} components alone. This is attributed to the
degeneracy between the mass of the lens and the external
convergence (Falco et al.~\cite{falco1985}; Gorenstein et
al.~\cite{gorenstein1988}; Bernstein \&
Fischer~\cite{bernstein1999}). The use of the time delay as an
additional constraint helped to distinguish between the mass of
the lens and the mass of the perturber, but there is still a
degeneracy between the perturber mass and its distance from the
lens. For the case when $z_{\mathrm{G_{p}}}>z_{\mathrm{l}}$ the
higher-redshift perturber is probably located at the smaller
impact parameter from the lensing galaxy (and also from the
quasar) and so affects its lensing properties.

More conclusions about the \object{UM673} lens model require
additional observations. In particular, the measurement of the
velocity dispersion of the lens galaxy will tightly constrain its
mass; the detailed analysis of the lens environment will help to
reduce errors in the lens model. A more accurate measurement of
the time delay between the \object{UM673} components will also
improve the mass model of the lens.

\begin{acknowledgements}

We thank the referee for useful comments and help in improving the
content of this paper. We would like to thank Luis Goicoechea for
interesting discussions on the topic. We thank the PhD student
H.H. Chan for his efforts in testing the NFW lens model based on
the data presented in this paper. We appreciate the Subaru time
allocation committee for scheduling our observations (proposal ID:
S12A0150S; PI: E. Koptelova). The research project was supported
by the National Taiwan University grant No. 10R40044 and by the
Taiwan National Science Councils grant No. NSC99-2811-M-002-051.
This research has made use of the ESO Science Archive data taken
with the VLT telescope.

\end{acknowledgements}

\end{document}